\documentclass[10pt,preprintnumbers,amsmath,amssymb,twocolumn,nofootinbib,superscriptaddress, floatfix]{revtex4-2}

\usepackage[english]{babel}

\usepackage[utf8x]{inputenc}
\usepackage[T1]{fontenc}
\usepackage{ucs}
\usepackage{microtype}
\usepackage{newtxtext,newtxmath}
\usepackage{color}
\usepackage{lettrine}

\usepackage{url}
\usepackage{hyperref}

\usepackage{xspace}
\usepackage{xcolor}
\usepackage{lastpage}
\usepackage{stmaryrd}

\usepackage{graphicx}
\usepackage{empheq}
\usepackage{ifthen}
\usepackage{tensor}
\usepackage{paralist}
\usepackage{float}

\usepackage{slashed}
\usepackage{amsmath}
\usepackage{amsfonts}
\usepackage{amssymb}
\usepackage{mathrsfs}
\usepackage{extarrows}
\usepackage{verbatim}
\usepackage{upgreek}
\usepackage{wasysym}
\usepackage{soul}

\usepackage{natbib}
\usepackage{ulem}

\def \be {\begin{equation}}
\def \ee {\end{equation}}
\def \dd {\mathrm{d}} 

\def \p {\partial}
\def \l {\left}
\def \r {\right}

\def \bs {\boldsymbol}

\def \S {\mathbb{S}}

\newcommand{\e}[1]{_{\rm #1}}

\newcommand{\beq}{\begin{equation}}
\newcommand{\eeq}{\end{equation}}
\newcommand{\bea}{\begin{eqnarray}}
\newcommand{\eea}{\end{eqnarray}}


\newcommand{\bn}{{\bf{n}}}

\newcommand\ees{\end{eqnarray}}
\newcommand\bees{\begin{eqnarray}}

\definecolor{magenta}{rgb}{0.1,0.98,0.6}
\definecolor{dgreen}{rgb}{0,0.7,0.0}

\begin{document}

\title{Statistical effects of the observer's peculiar velocity on source number counts}

\author{Charles Dalang}
\email{charles.dalang@gmail.com}
\affiliation{Universit\'e de Gen\`eve, D\'epartement de Physique Th\'eorique and Center for Astroparticle Physics, 24 quai Ernest-Ansermet, CH-1211 Gen\`eve 4, Switzerland}
\affiliation{Queen Mary University of London, Mile End Road, London E1 4NS, United Kingdom}
\author{Ruth Durrer}
\affiliation{Universit\'e de Gen\`eve, D\'epartement de Physique Th\'eorique and Center for Astroparticle Physics, 24 quai
 Ernest-Ansermet, CH-1211 Gen\`eve 4, Switzerland}
\author{Fabien Lacasa}
\affiliation{Universit\'e de Gen\`eve, D\'epartement de Physique Th\'eorique and Center for Astroparticle Physics, 24 quai Ernest-Ansermet, CH-1211 Gen\`eve 4, Switzerland}
\affiliation{Dipartimento di Fisica e Astronomia ``Galileo Galilei'', Università degli Studi di Padova, via Marzolo 8, I-35131, Padova, Italy.}
\affiliation{INFN, Sezione di Padova, via Marzolo 8, I-35131, Padova, Italy.} 
\affiliation{Université Paris-Saclay, CNRS, Institut d’Astrophysique Spatiale, 91405, Orsay, France}

\begin{abstract}
The velocity of the Sun with respect to the cosmic microwave background (CMB) can be extracted from the CMB dipole, provided its intrinsic dipole is assumed to be small in comparison. This interpretation is consistent, within fairly large error bars, with the measurement of the correlations between neighboring CMB multipoles induced by the velocity of the observer, which effectively breaks isotropy. In contrast, the source number count dipole was reported to privilege a velocity of the observer with an amplitude which is about twice as large as the one extracted from the entirely kinematic interpretation of the CMB dipole,  with error bars which indicate a more and more significant tension. In this work, we study the effect of the peculiar velocity of the observer on correlations of nearby multipoles in the source number counts. We provide an unbiased estimator for the kinematic dipole amplitude, which is proportional to the peculiar velocity of the observer and we compute the expected signal to noise ratio. Assuming full sky coverage, near future experiments can achieve better than 5$\%$ constraints on the velocity of the Sun with our estimator.
\end{abstract}

\maketitle


\section{Introduction}\label{sec:intro}
\begin{flushleft}
\textit{Most humans have two points of view,  which are as close as their eyes are, giving them only a little bit of perspective. \\
\indent   In this sense, hammerhead sharks have more.}
\end{flushleft}

\lettrine{W}{atching} the stars on a clear night sky may easily make one wonder about one's place in the Universe. From such a point of view, it seems hard not to think that one is located in a very special place. Be that as it may, the \textit{Copernican principle} states that humans are not privileged observers of the Universe. Combined with evidence of statistical isotropy in the temperature associated with a black body spectrum from the Cosmic Microwave Background (CMB) up to temperature fluctuations of the order of $10^{-5}$ \cite{Aghanim:2018eyx}, the Copernican principle points towards the \textit{cosmological principle}, which is a corner stone of modern cosmology: The Universe is statistically homogeneous and isotropic. The implementation of these strong assumptions allows for the use of the highly symmetric Friedmann-Lemaître-Robertson-Walker (FLRW) metric and the application of perturbation theory to describe the Universe, on sufficiently large scales. This drastically  simplifies calculations and has allowed us to constrain with percent precision the $6$-parameter standard model of cosmology, the $\Lambda$-Cold-Dark-Matter ($\Lambda$CDM) model~\cite{Aghanim:2018eyx}.

In this reasoning, one important step was swept under the carpet. To recover statistical isotropy of CMB temperature fluctuations of the order of $10^{-5}$, one must boost the observer to the so-called \textit{CMB frame} in which the apparently large CMB dipole, which is of order $10^{-3}$, i.e.\,a hundred times larger than other multipoles, vanishes \cite{Fixsen:1994, Fixsen:1996,Aghanim:2018eyx,Planck:2013kqc}. This is called the \textit{entirely kinematic interpretation} of the CMB dipole and is motivated by standard single field inflation. This canonical model predicts a nearly scale invariant power spectrum of primordial fluctuations of the inflaton field, at the end of a period of quasi de Sitter expansion. In this context, there is no reason to expect the dipole to be a hundred times larger than other multipoles and one rather expects a primordial intrinsic dipole of the order of $10^{-5}$. On the other hand, one also expects a CMB dipole to be generated by the peculiar velocity of the observer, as pointed out by D.W.\,Sciama \cite{Sciama:1967} and calculated by P.J.E.\,Peebles and D.T.\,Wilkinson \cite{Peebles:1968}. One can easily be tempted to attribute the large dipole to the velocity of the observer with respect to the CMB, potentially absorbing a small intrinsic dipole, which is expected to yield a $1\%$ correction. Absorbing the entire CMB dipole in the velocity leads to an observer velocity $||\bs{v}_o|| = 369.82 \pm 0.11$ km s$^{-1}$, corresponding to $\beta\e{dip}=||\bs{v}_o||/c=0.00123\pm 0.00036$, pointing towards $\bs{\hat{\beta}}\e{dip}=(264.021^\circ\pm 0.011^\circ,48.253^\circ\pm 0.005^\circ)$ in galactic coordinates, where the yearly modulation of $\sim 30$ km s$^{-1}$ of the Satellite orbiting around the Sun has been removed \cite{Aghanim:2018eyx}. While this order of magnitude for the peculiar velocity of the Sun is expected, it is important to recall that an intrinsic CMB dipole is degenerate with a boost to linear order in $\beta$. This implies that a boost to the CMB frame, may not correspond to the frame in which matter fields live at rest on average, if there exists a significant intrinsic cosmic dipole \cite{Domenech:2022mvt,Agullo:2020fbw}. This may potentially have dramatic consequences on the interpretation of the CMB or of Supernova data. (See for example \cite{Freeman:2005nx,Naselsky_2012,Colin:2019opb,Mohayaee:2021jzi}). 

There are several alternative ways to measure the velocity of the observer with respect to the rest frame of matter fields. One can check that the distribution of far enough sources\footnote{Far enough sources means sources with observed redshift $z\geq 0.1$, such that the intrinsic dipole, predicted in $\Lambda$CDM is small in comparison to the kinematic dipole  \cite{Tiwari:2016,Dalang:2021ruy}.} leads to a kinematic dipole $\mathcal{D}\e{kin}$, consistent with the CMB dipole. This was pioneered by G.\,Ellis and J.\,Baldwin \cite{Ellis1984}, who determined  the dipole in the source number counts per unit solid angle for radio sources with a flux following a power law spectrum in frequency. Different teams reported a source number count dipole with a direction which agrees fairly well with the CMB dipole, but with an amplitude which is about twice as large as expected from the CMB dipole \cite{Secrest:2020has, Secrest:2022uvx,Bengaly:2017slg,Siewert:2021}. It was suggested by \cite{Dalang:2021ruy} that evolution bias may, at least partially, explain the reported tension. This was further studied by the authors of \cite{Guandalin:2022tyl}, which also find significant variations in the number count kinematic dipole in the presence of parameter evolution when using different quasar luminosity function models. Combining mid-infrared quasars and radio sources, a $5.2\sigma$ tension between these two dipoles was reported in \cite{Secrest:2022uvx}. The authors of \cite{Dam:2022wwh} reanalyzed the same mid-infrared quasars and concluded that neither masking, nor parameter evolution could fully explain the reported tension, even if this is subject to further assumptions. Let us, however, also mention that a source number count dipole performed with data from the Very Large Array Sky Survey and the Rapid Australian Square Kilometer Array Pathfinder Continuum Survey was reported to be consistent with the CMB dipole \cite{Darling:2022jxt}, although with much larger error bars. Note that dipole measurement with $1048$ supernovae type 1a from the Pantheon sample was found to point in a direction well aligned with the CMB dipole with an amplitude which is $2.4\sigma$ smaller than the CMB dipole \cite{Horstmann:2021jjg}. Carefully analyzing the peculiar velocity field with different catalogues, the authors of \cite{Ma:2011} found a bulk flow velocity of about $400$ km s$^{-1}$ extending to $\lesssim 150 h^{-1}$Mpc, which is difficult to accommodate within the standard cosmological paradigm. The cosmic infrared background can also be used to measure the velocity of far away galaxies relative to the observer and a signal to noise ratio of $50-100$ is forecasted for the Euclid and Roman surveys \cite{Kashlinsky:2022tit}. Gravitational waves also offer promising ways to measure the kinematic dipole \cite{Mastrogiovanni:2022nya,Chung:2022xhv}. Additionally, we also point out that the degeneracy between intrinsic and kinematic dipole may be broken using the redshift dependent dipoles in future source number count experiments \cite{Nadolny:2021hti}. Fluctuations of the number counts from radio surveys such as the Square Kilometer Array may also be useful to constrain the observer's peculiar velocity \cite{Pant:2018smd}.

The CMB, however, appears to be consistent with itself. Indeed, the velocity of the observer which, effectively breaks statistical isotropy, induces correlation of the $l$ and $(l\pm1)$ multipoles in the CMB~\cite{Challinor:2002zh,Dai:2014,Chluba:2011}. This leads to an independent measurement of $\beta$, which was found to be $\beta =0.00128\pm 0.00026(\hbox{stat.})\pm 0.00038(\hbox{syst.})$ in \cite{Planck:2013kqc}, consistent with $\beta\e{dip}$. Slightly tighter constraints, consistent with $\beta\e{dip}=0.00123$ were obtained in the analysis presented in \cite{Saha:2021bay,Ferreira:2020aqa}. This shows that the entirely (or at least dominantly) kinematic interpretation of the CMB dipole is consistent with the correlation of neighboring multipoles, although there is still room for an intrinsic CMB dipole, which can make up a significant portion of the observed CMB dipole, without contradicting the observed $l$ and $(l\pm1)$ correlations \cite{Schwarz:2016}. However, a significant bulk flow of galaxy clusters extending up to $300 h^{-1}$Mpc was found in WMAP via the kinematic Sunyaev-Zeldovich effect \cite{Kashlinsky:2008ut}.  Let us also mention that while a boost and an intrinsic dipole are degenerate in the CMB dipole at linear order in $\beta$, second order corrections in $\beta$ give distinct spectral distortions in the CMB monopole and quadrupole, which may allow for a measurement of the intrinsic CMB dipole in futuristic CMB spectral distortions experiments, as pointed out in~\cite{Yasini:2016dnd}. 

In this work, we study the correlation of neighboring multipoles in the source number counts, which yield an independent crosscheck of the validity of the measurement of the kinematic dipole 
with source number counts. We find that the peculiar velocity of the observer induces correlations between the $l$ and $(l\pm 1)$ multipoles in the  source number counts, which are absent for comoving observers in a statistically 
isotropic Universe. Assuming full sky coverage, we derive an unbiased estimator $\hat{\mathcal{D}}\e{kin}$ of the kinematic dipole amplitude, proportional itself to $\beta$, and compute its variance and signal to noise ratio. We also comment on the determination of the velocity direction and on limitations from a partial sky survey.

The paper is structured as follows; In Sec.\,\ref{sec:Setup}, we present the setup and the transformation rules of the ingredient quantities under  boost. In Sec.\,\ref{sec:Source_Number}, we detail the computation of the boosted spherical harmonic coefficients of the source number counts for redshift surveys and redshit independent ones. In Sec.\,\ref{sec:Correlation}, we calculate the 2-point correlation function for an observer boosted in an otherwise isotropic Universe. In Sec.\,\ref{sec:Estimator}, we write down an estimator for the amplitude of the dipole and for the velocity of the observer with respect to the rest frame of distance sources. We show that this estimator is unbiased and impose an upper bound for its variance in terms of measured quantities. In Sec.\,\ref{sec:orientation}, we briefly outline the determination of the orientation of the peculiar velocity. We discuss our findings, the potential for a measurement and its limitations in Sec.\,\ref{sec:Discussion}.\\ Units are such that $c=1$, bold symbols indicate three dimensional vectors, hats may indicate unit vectors when used on bold symbols or statistical estimators.


\section{The Setup}\label{sec:Setup}

We consider two observers $O$ and $O'$ which are related by a Lorentz boost of velocity $\beta = ||\bs{v}_o||/c$. Here $\bs{v}_o$ is the velocity of $O'$ with respect to $O$, which is aligned with their respective $\bs{\hat{z}} = \bs{\hat{z}'}$ axis. This choice allows for the azimuthal  angles of the two observers to coincide, $\varphi = \varphi'$, despite the Lorentz transformation. In Sec.\,\ref{sec:orientation}, we consider arbitrary directions of the peculiar velocity of the observer. Primed quantities relate to $O'$ and quantities without primes relate to $O$. We assume that both of these observers live in an FLRW Universe described by the  line element $\dd s^2 = a^2(\eta)(-\dd \eta^2 + \dd r^2 + r^2 (\dd \theta^2 + \sin^2\theta \dd \varphi^2 ))$, where $a(\eta)$ is the scale factor, which only depends on conformal time $\eta$ and $r, \theta$ and $\varphi$ are comoving coordinates. Observer $O$ is assumed to be following the Hubble flow, i.e.\,having zero peculiar velocity and may be called a \textit{comoving} observer. The motion of observer $O'$ with respect to $O$ affects their measurements of time intervals, cosines of polar angles, polar angles, solid angles, frequencies and redshift, which transform respectively, in the following way
\begin{align}
\dd t' & = \dd t \sqrt{1- \beta^2}\,, \label{eq:TR1}\\
\cos \theta' & = (\cos \theta + \beta)[1 + \beta \cos \theta]^{-1}\,, \label{eq:TR2}\\
\theta' & =  \theta + \delta\theta, \qquad 
\delta\theta = -\beta\sin\theta + {\cal O}(\beta^2)\,,\\
\dd \Omega' & = \dd^2\bs{\hat{n}'}= \sin(\theta')\dd \theta' \dd \varphi'= \dd \Omega (1- \beta^2) [1 + \beta \cos \theta]^{-2}\,, \label{eq:TR3}\\
\nu' & = \nu (1+\beta  \cos \theta)\gamma \,,\label{eq:TR4}\\
1+z' &=(1+z)(1+\beta  \cos \theta)^{-1}\gamma^{-1}\,, \label{eq:TR5}
\end{align}
where $z$ and $z'$ denote the redshifts of a photon observed at an angle $\theta$ (respectively $\theta'$) with respect to $\bs{v}_o$ and $\gamma = 1/\sqrt{1-\beta^2}$ is the Lorentz factor. Observer $O'$ calls $\bs{\hat{n}'}$ the direction of an incoming photon which corresponds to $\bs{\hat{n}}$ for observer $O$. Those are related by~\cite{Jackson:1975}
\begin{align}
\bs{\hat{n}'} = \l( \frac{\bs{\hat{n}} \cdot \bs{\hat{\beta}} + \beta }{1 + \bs{\hat{n}} \cdot \bs{\beta}}\r) \bs{\hat{ \beta}} + \frac{\bs{\hat{n}} - \l(\bs{\hat{n}} \cdot \bs{\hat{\beta}}\r) \bs{\hat{\beta}}}{\gamma (1+\bs{\hat{n}}\cdot \bs{\beta})}\,.\label{eq:np_to_n}
\end{align}
In the next section, we apply these transformation rules to the source number count.


\section{Source Number counts}\label{sec:Source_Number}
We express the number of sources $\dd N$ per unit solid angle $\dd \Omega$ and per redshift bin $\dd z$ in the direction $\bs{\hat{n}}\, \widehat{=}\, (\theta,\varphi)$ and at redshift $z$ for sources with  flux  $S$ (in W m$^{-2}$ Hz$^{-1}$) above a certain  treshold $S_*(\nu_o)$ in some frequency band $[\nu_o, \nu_o +\dd \nu_o]$ as
\begin{align}
\frac{\dd N}{\dd \Omega \dd z} [\bs{\hat{n}},z, S>S_*(\nu_o)]\,.
\end{align} 
For fixed $S_*(\nu_o)$ and fixed redshift $z$, the real-valued function $\frac{\dd N}{\dd \Omega\dd z}: \mathbb{S}_2\to \mathbb{R}$ may be expanded in complex spherical harmonics
\begin{align}\label{eq:Spherical_Harmonics_Expansion}
\frac{\dd N}{\dd \Omega\dd z} [\bs{\hat{n}},z, S>S_*(\nu_o)]  = \sum_{l=0}^{+\infty} \sum_{m=-l}^{l} a_{l m}(z) Y_{l m}(\bs{\hat{n}})\,.
\end{align}
with complex valued functions $a_{lm}(z):\mathbb{R}_+ \to \mathbb{C}$, with $l\in \mathbb{N}$ and $m \in [-l,-l+1,\dots, l-1,l]$. In case, the survey lacks redshift information, we also consider the same quantity integrated over redshift, which we denote by
\begin{align}
\frac{\dd N}{\dd \Omega }[\bs{\hat{n}}, S>S_*(\nu_o)] & \equiv \int_0^{+\infty} \dd z \,\frac{\dd N}{\dd \Omega \dd z} [\bs{\hat{n}},z, S>S_*(\nu_o)]\nonumber \\
& = \sum_{l=0}^{+\infty} \sum_{m=-l}^{l} a_{l m} Y_{l m}(\bs{\hat{n}})\,, \label{eq:NumberCountSphericalHarmonics}
\end{align}
and for which the expansion coefficients are redshift independent. Using the orthogonality relations of the spherical harmonics
\begin{align}\label{eq:orthogonality_relations}
\int_{\S_2}\dd^2\bs{\hat{n}}\,Y_{lm}(\bs{\hat{n}}) Y_{l'm'}^*(\bs{\hat{n}}) = \delta_{ll'}\delta_{mm'}\,,
\end{align}
one can obtain the expansion coefficients,
\begin{align}
a_{lm}(z) = \int_{\S_2} \dd^2\bs{\hat{n}}\, \frac{\dd N}{\dd \Omega \dd z}[\bs{\hat{n}},z,S>S_*(\nu_o)] Y^*_{lm}(\bs{\hat{n}})\,.
\end{align}
The redshift independent coefficients are obtained in a similar way, 
\begin{align}
a_{lm} = \int_{\S_2} \dd^2\bs{\hat{n}}\, \frac{\dd N}{\dd \Omega }[\bs{\hat{n}},S>S_*(\nu_o)] Y^*_{lm}(\bs{\hat{n}})\,.
\end{align}
A masked sky breaks these orthogonality relations, of course. In the following, we distinguish between an isotropic Universe and a \textit{statistically} isotropic Universe. In an isotropic Universe, only the monopole, i.e.\,$a_{00}$ contributes to the sum in Eq.\,\eqref{eq:Spherical_Harmonics_Expansion}. This corresponds to the case where the number density of sources is exactly the same in every direction. In practice, isotropy is only true in a statistical sense.
The expectation value of the number of galaxies is the same in every direction.  Within the standard model, gravity acts to cluster sources, such that in a statistically isotropic Universe, there are in principle an infinite series of $a_{lm}$'s which contribute. The $a_{lm}$'s of higher $l$  correspond to perturbations on smaller and smaller angular scale. So far, number counts have been analysed in angular space mainly in photometric surveys which make a 3x2pt analysis of shear, number counts and their cross-correlations, see e.g.~\cite{DES:2021bwg,KiDS:2020ghu}. A harmonic space analysis of the number counts from the Dark Energy Survey (DES) data is found e.g. in~\cite{DES:2018csk,DES:2021dpy}. Importantly, the motion of the observer in an otherwise isotropic Universe contributes a dipole to the source number count such that the number of sources per unit solid angle and redshift observed by $O'$ at redshift $z'$ in direction $\bs{\hat{n}'}$ reads \cite{Maartens:2017qoa, Dalang:2021ruy} 
\begin{align}\label{eq:Isotropic_Dipole}
\frac{\dd N'}{\dd \Omega'\dd z'}[\bs{\hat{n}}',z', & S>S_*(\nu_o)] =   \Big( 1 + \left(\bs{\hat\beta }\cdot \bs{\hat{n}}\right) \, \mathcal{D}\e{kin}(z) \Big) \\
& \times \l \langle \frac{\dd N}{\dd \Omega\dd z}[z,\bs{\hat{n}}, S>S_*(\nu_o)]\r \rangle + \mathcal{O}(\beta^2)\,,
\end{align}
where the amplitude of the dipole is given by
\begin{align}\label{eq:Redshift_Dipole}
\mathcal{D}\e{kin}(z) = \l[ 2 + \frac{2[1-x(z)]}{r(z)\mathcal{H}(z)} + \frac{\dot{\mathcal{H}}(z)}{\mathcal{H}^2(z)} - f\e{evol}(z)\r] \beta \,.
\end{align}
Here $\mathcal{H}(z)= \dot{a}(\eta)/a(\eta)$ indicates the conformal Hubble rate, a dot indicates a derivative with respect to conformal time and $r(z)$ is the background comoving distance,
\begin{align}
r(z)=  \int_0^{z} \frac{\dd z}{(1+z)\mathcal{H}(z)}\,. 
\end{align}
The magnification bias $x(z)$ (sometimes noted $s(z) = 2 x(z)/5$) is defined as
\begin{align}
x(z) \equiv -\frac{\p \ln  \l \langle \frac{\dd N}{\dd \Omega \dd z}[\bs{\hat{n}},z, S>S_*]\r \rangle }{\p \ln S_*}\,,\label{eq:x_z}
\end{align}
and is sometimes defined in its integrated form $\l \langle \frac{\dd N}{\dd \Omega \dd z}[\bs{\hat{n}},z, S>S_*]\r \rangle \propto S_*^{-x(z)}$. It tracks the number density of sources above a given flux density treshhold $S_*$. Intuitively, for positive $x(z)$, the number density of objects above the treshold $S_*$ goes to zero for large enough $S_*$ and diverges for $S_*$ going to zero. A constant $x(z)$ means that this power law does not change with redshift or, alternatively that the population distribution of fluxes is constant in cosmic time. The evolution bias traces the time evolution of the number of sources per unit comoving volume $\dd V$
\begin{align}
f\e{evol}(z) & \equiv \frac{1}{\mathcal{H}}  \l \langle \frac{\dd N}{\dd V}[r(z), L>L_*] \r \rangle^{-1} \frac{\p}{\p \eta}  \l \langle \frac{\dd N}{\dd V}[r(z), L>L_*] \r \rangle \nonumber \\
& = - \frac{\p \ln \l \langle \frac{\dd N}{\dd V}[r(z), L>L_*] \r \rangle }{\p \ln (1+z)}\,.
\end{align}
Here $L$ and $L_*$ are luminosity densities (in W Hz$^{-1}$) corresponding to the flux densities $S$ and $S_*$, respectively. The angular brackets\footnote{Later in the manuscript, angular brackets denote expectation values. The context should allow the experienced reader to break this degeneracy.} indicate an average over the 2-sphere $\S_2$
\begin{align}
\l \langle \dots \r \rangle  = \frac{1}{4\pi} \int_{\S_2} \dots \,\dd^2\bs{\hat{n}} \,.
\end{align}
The difference between $\bs{\hat{n}}'$ and $\bs{\hat{n}}$ results in second order corrections (i.e.\,$\mathcal{O}(\bs{\beta}^2)$) of $\dd N'/(\dd \Omega'\dd z')[\bs{\hat{n}'},S>S_*(\nu_o)]$ in a Universe where the sources are isotropically distributed. Instead, in a Universe which is \textit{statistically} isotropic, the difference between $\bs{\hat{n}}$ and $\bs{\hat{n}'}$ becomes first order in $\beta$. We note $\bs{\hat{n}'} \, \widehat{=}\, (\theta', \varphi')=(\theta+\delta \theta, \varphi)$. Recall that since we have assumed that $\bs{\beta}$ is aligned with $\bs{\hat{z}}$, the azimutal angle $\varphi$ is unaffected by the boost. In a Universe with intrinsic anisotropies, an additional term in the number count plays a role. More precisely, Eq.\,(25) in \cite{Dalang:2021ruy} becomes 
\begin{widetext}
\begin{align}
\frac{\dd N'}{\dd \Omega' \dd z'} [\bs{\hat{n}}',z', S>S_*] & =\frac{\dd N'}{\dd \Omega' \dd z'} ( \theta +\delta \theta, \varphi, r[z',\bs{\hat{n}}], L> L_*'[z',\bs{\hat{n}}, \nu_s]) \\
& \simeq \frac{\dd N}{\dd \Omega \dd z} [ \bs{\hat{n}}, r[z], L>L_*] \left(\frac{\dd \Omega}{\dd \Omega'}  +\frac{\dd z}{\dd z'}\right) \nonumber\\
& \quad + \frac{\p}{\p r'} \l( \frac{\dd N}{\dd \Omega \dd z}[\bs{\hat{n}}, r',L>L_*]\r) \Big |_{r'=r[z]}\cdot \delta r[z,\bs{\hat{n}}] \nonumber\\
& \quad + \frac{\p}{\p L_*'}\l( \frac{\dd N}{\dd \Omega \dd z}[\bs{\hat{n}}, r,L>L'_*]\r) \Big|_{L_*' =L_*} \cdot \delta L_*[z,\bs{\hat{n}},\nu_s] \nonumber\\
& \quad + \frac{\p}{\p \theta'} \l( \frac{\dd N}{\dd \Omega \dd z}[\theta', \varphi,r, L>L_*]\r)\Big|_{ \theta' = \theta} \cdot \delta \theta [\bs{\hat{n}}]
\end{align}
\end{widetext}
where in the first line, we have rewritten $\bs{\hat{n}'}
\, \widehat{=}\, (\theta', \varphi') = (\theta + \delta \theta, \varphi)$. We associated a direction dependent comoving distance $r[z',\bs{\hat{n}}] = r[z'] + \delta r[z',\bs{\hat{n}}]$ to sources located at fixed observed redshift (with $\delta r[z,\bs{\hat{n}}] = -\bs{\hat{n}}\cdot \bs{\beta}/\mathcal{H}$). We also associated a direction dependent luminosity density treshold $L_*'[z',\bs{\hat{n}}, \nu_s']= L_*[z',\nu_s']+ \delta L_*[z',\bs{\hat{n}},\nu_s']$, which corresponds to a fixed observed flux density treshold of the detector, which is independent of its motion. In the second line, we have changed the "\textit{per observed}" unit solid angle $\dd \Omega'$ and redshift interval $\dd z'$ to "\textit{per background}" unit solid angle $\dd \Omega$ and redshift $\dd z$, according to Eqs.\eqref{eq:TR1}-\eqref{eq:TR5}. In the following 3 lines, we Taylor expand around background quantities the three variables which are affected by the boost, namely, $r$, $L_*$ and $\theta$. The only term which was not accounted for in \cite{Dalang:2021ruy}, is the $\p_{\theta'}$ derivative. The $\p_{\theta'}$ derivative is equal to zero to first order in $\beta$ in case the number counts are independent of angular direction for observer $O$. This is equivalent to assuming that the number counts are isotropic for observer $O$ and consist strictly of a monopole, which is an assumption of \cite{Dalang:2021ruy} and is sufficient if one is only interested in the dipole generated by the monopole due to the motion of the observer. Here, however we want to study the modification of all multipoles due to the motion of the observer. This is because $\delta \theta(\bs{\hat{n}})=-\beta \sin \theta$ is linear in $\beta$ and  the $\p_{\theta'}$ derivative on a monopole which, by definition is independent of angles, vanishes. However, if there are intrinsic anisotropies, meaning if $a_{lm}\neq 0$ for $l\geq 1$, this derivative is non zero. We have $\delta \theta = -\beta \sin\theta = - \tan(\theta) \bs{\beta} \cdot \bs{\hat{n}}$. 
This $\p_{\theta'}$ derivative results in an additional dipolar term which comes from relating the number density from direction $\bs{\hat{n}'}$ for the boosted observer to direction $\bs{\hat{n}}$ for the observer at rest with respect to the source's rest frame. Therefore, instead of Eq.\,\eqref{eq:Isotropic_Dipole}, in a statistically isotropic Universe, we get
\begin{align}\label{eq:Boosted_number_counts_z}
\frac{\dd N'}{ \dd \Omega' \dd z'}[\bs{\hat{n}'},z',S>S_*] & =\l( 1 +  \l[\cos\theta\,\mathcal{D}\e{kin}(z) - \beta\sin (\theta)\p_\theta \r] \r) \nonumber \\
& \quad \times \frac{\dd N}{\dd \Omega \dd z} [\bs{\hat{n}},z,S>S_*]\,.
\end{align}
Assuming that the sources have a luminosity density which follows a frequency power law $L \propto \nu_s^{-\alpha(z)}$ with spectral index $\alpha(z)$, one can integrate over $\dd z$, write the partial derive of $f\e{evol}(z)$ in terms of a total redshift derivative, relate the luminostiy density to a flux density and integrate by parts (see Sec.\,II of \cite{Dalang:2021ruy} or App.\,A of \cite{Nadolny:2021hti}) to find
\begin{widetext}
\begin{align}\label{eq:Integrated_Step1}
\frac{\dd N'}{ \dd \Omega' }[\bs{\hat{n}'},S>S_*] = \int_0^{+\infty} \dd z  \l( 1 + \bs{\hat{n}}\cdot \bs{\beta} \l[3 + x(z) [1 + \alpha(z)] - \tan (\theta)\p_\theta  + (1+z)\frac{\dd }{\dd z}\r] \r)  \frac{\dd N}{\dd \Omega \dd z}  [\bs{\hat{n}},z,S>S_*] \,.
\end{align}
Assuming for simplicity that $x(z)$ and $\alpha(z)$ are constant, we find, after integrating by parts and neglecting boundary terms\footnote{Boundary terms may not necessarily vanish. For example, if one works with redshift bins, one may have to include these boundary terms, which are straightforward to compute from Eq.\,\eqref{eq:Integrated_Step1}.} that
\begin{align}\label{eq:Boosted_number_counts}
\frac{\dd N'}{ \dd \Omega' }[\bs{\hat{n}'},S>S_*] =  \Big( 1 + \l[\cos\theta\,\mathcal{D}\e{kin}
- \beta\sin (\theta)\p_\theta \r]  \Big) \frac{\dd N}{\dd \Omega}[\bs{\hat{n}},S>S_*]\,,
\end{align}
\end{widetext}
where the kinematic dipole boils down to the Ellis and Baldwin formula\footnote{Stricktly speaking, it is sufficient that $\alpha(x)$ and $x(z)$ are uncorrelated to recover the Ellis and Baldwin formula.} \cite{Ellis1984}
\begin{align}\label{eq:Ellis_Baldwin}
\mathcal{D}\e{kin} = [2 + x(1+\alpha)] \beta \,.
\end{align}
Recall that a constant $x$ means that the flux distribution of sources does not depend on redshift, as we have discussed below Eq.\,\eqref{eq:x_z}. Typical values of these parameters for radio galaxies are $x\sim 1$ and $\alpha \sim 1$, such that $\mathcal{D}\e{kin}\sim 4 \beta$.  From now on, we focus on the redshift integrated surveys, which measure $\dd N' /\dd \Omega'$, but  comparing \eqref{eq:Boosted_number_counts_z} and \eqref{eq:Boosted_number_counts}, one sees that the only change in  redshift dependent surveys is the  change of $\mathcal{D}\e{kin}$ in \eqref{eq:Boosted_number_counts} to $\mathcal{D}\e{kin}(z)$, defined in \eqref{eq:Redshift_Dipole}.

For observer $O'$, the number of sources per unit solid angle may also be expanded in spherical harmonics, although the coefficients will in general be different
\begin{align}
\frac{\dd N'}{ \dd \Omega'}[\bs{\hat{n}'}, S>S_*(\nu_o)]  = \sum_{l=0}^{+\infty} \sum_{m=-l}^l a'_{lm}Y_{lm}(\bs{\hat{n}'}) \,.
\end{align}
One computes the boosted $a'_{lm}$'s by computing the following integrals
\begin{align}\label{eq:boosted_alm_measured}
a_{lm}' = \int_{\S_2} \dd^2 \bs{\hat{n}'} \, \frac{\dd N'}{ \dd \Omega'}[\bs{\hat{n}'}, S>S_*(\nu_o)]  Y^*_{lm}(\bs{\hat{n}'})\,.
\end{align}
One can express these integrals as follows
\begin{widetext}
\begin{align}
a'_{l' m'} & = \int_{\S_2} \dd^2 \bs{\hat{n}'}\, \frac{\dd N'}{\dd \Omega'}[\bs{\hat{n}'}, S>S_*(\nu_o)] Y_{l' m'}^*(\bs{\hat{n}'}) \nonumber \\
& = \int_{\S_2} \dd^2 \bs{\hat{n}'} \, \l[  \Big( 1 + \l[\cos\theta\,\mathcal{D}\e{kin} - \beta\sin (\theta)\p_\theta \r]  \Big)  \frac{\dd N}{\dd \Omega} [\bs{\hat{n}}, S>S_*(\nu_o)]\r] \cdot Y_{l' m'}^*(\bs{\hat{n}'}) \nonumber \\
& = \int_{\S_2} \dd^2 \bs{\hat{n}'} \, \l[  \Big( 1 + \l[\cos\theta\,\mathcal{D}\e{kin} - \beta\sin (\theta)\p_\theta \r]  \Big)   \sum_{l m} a_{l m} Y_{l m}(\bs{\hat{n}})\r] \cdot (-1)^{m'} Y_{l' -m'}(\bs{\hat{n}'})\,, \label{eq:11}
\end{align}
\end{widetext}
where we have used that $Y^*_{l m}(\bs{\hat{n}'}) =(-1)^m Y_{l (-m)}(\bs{\hat{n}'})$ and expanded the number counts in spherical harmonics according to Eq.\,\eqref{eq:NumberCountSphericalHarmonics}. We express $Y_{lm}(\bs{\hat{n}})$ in terms of the variable $\bs{\hat{n}'}$ by a Taylor expansion
\begin{align}
Y_{lm}(\bs{\hat{n}}) = Y_{lm} (\bs{\hat{n}'}) + \beta \sin \theta \p_\theta Y_{lm} (\bs{\hat{n}'}) + \mathcal{O}(\beta^2 )\,.
\end{align}
This cancels the $-\beta \sin \p_\theta$ acting on the spherical harmonic $Y_{lm}(\bs{\hat{n}})$ in Eq.\eqref{eq:11} to first order in $\beta$. Using
\begin{align}
\cos \theta= 2 \sqrt{\frac{\pi}{3}}Y_{10}(\bs{\hat{n}})  \,,
\end{align}
we are left with
\begin{align}
a'_{l' m'}  & = a_{l' m'} \\
& +2\sqrt{\frac{\pi}{3}}\mathcal{D}\e{kin}(-1)^{m'}\sum_{lm}\int_{\S_2} \dd^2\bs{\hat{n}}\,Y_{10}(\bs{\hat{n}}) Y_{l m}(\bs{\hat{n}}) Y_{l' -m'}(\bs{\hat{n}})\,.\nonumber\\
\label{e:almprime}
\end{align}

\begin{widetext}
\begin{figure}
    \centering
    \includegraphics[width=2 \columnwidth]{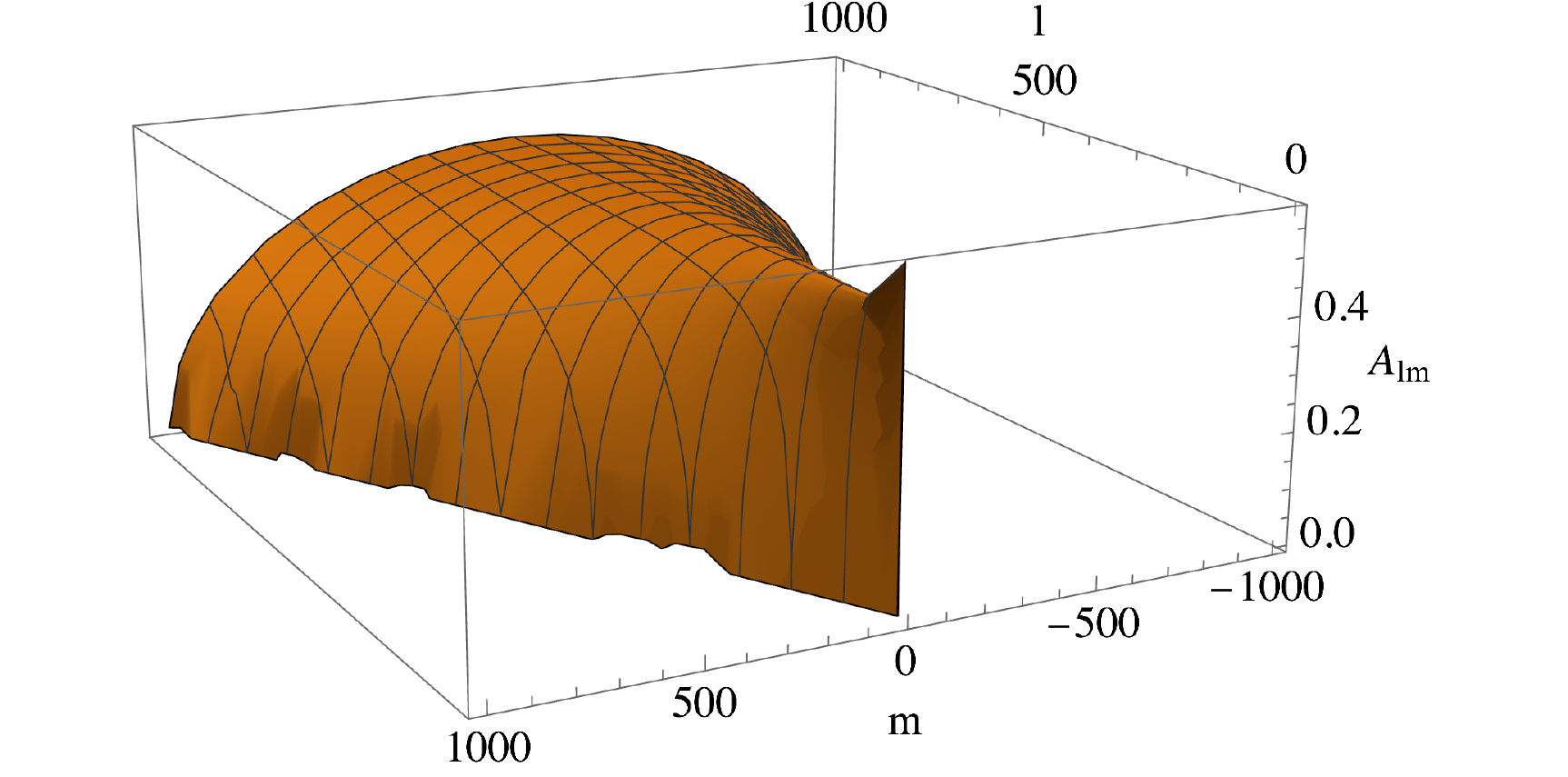}
    \caption{We plot the coefficients $A_{lm}$ as a function of $l\in[1,1000]$ and $m\in [-l,l]$. The $A_{lm}$'s  appear as a proportionality factors in front of $\mathcal{D}\e{kin}$ in the correlation of the neighboring multipoles in Eqs.\,\eqref{eq:Dkin_lplus1}. For $m=|l|$, they vanish. For $m< |l|$, $A_{lm}>0$, which allows for the use of the off diagonal multipole correlations to constrain the kinematic dipole $\mathcal{D}\e{kin}$.}
    \label{fig:A}
\end{figure}
\end{widetext}

The integrals involving three spherical harmonics are Gaunt coefficients, which satisfy certain selection rules. In particular, the three $l$'s which must satisfy the triangle condition, i.e.\,$|l_1-l_2|\leq l_3 \leq (l_1 + l_2)$. Then they are given by (see Appendix 4 of \cite{Durrer:2020fza} for more details)
\begin{align}
&\int_{\S_2} \sin \theta \dd \theta \dd \varphi \, Y_{l_1m_1}(\theta,\varphi)Y_{l_2m_2}(\theta,\varphi)Y_{l_3m_3}(\theta,\varphi) \\
& = \sqrt{\frac{(2l_1 +1)(2l_2 + 1)(2 l_3+1)}{4 \pi}} \begin{pmatrix} l_1 & l_2 & l_3 \\ 0 & 0 & 0 \end{pmatrix}\begin{pmatrix} l_1 & l_2 & l_3 \\ m_1 & m_2 & m_3 \end{pmatrix}\,, \nonumber 
\end{align}
else, these integrals vanish. The ($3\times2$) matrices are 3-j symbols which are related to the Clebsch-Gordan coefficients. They are non-vanishing only if the sum $m_1+m_2+m_3$ vanishes and the triangle inequality between the $l_i$'s is satisfied, see~\cite{Durrer:2020fza} or some text on quantum mechanics. Since in \eqref{e:almprime} there is a sum over $l$ and $m$ and since the triangle condition must hold, we have schematically,
\begin{align}
\sum_{l m}&\int_{\S_2} \sin \theta \dd \theta \dd \varphi \,Y_{l m}(\theta,\varphi)Y_{1 0}(\theta,\varphi)Y_{l' (-m')}(\theta,\varphi) \\
& = \sqrt{\frac{(2l'+3)3 (2 l'+1)}{4 \pi}} \begin{pmatrix} l'+1 & 1 & l'\\ 0 & 0 & 0 \end{pmatrix}\begin{pmatrix} l'+1 & 1 & l' \\ m' & 0 & -m' \end{pmatrix} \nonumber\\
& + \sqrt{\frac{(2l'-1))3 (2 l'+1)}{4 \pi}} \begin{pmatrix} l'-1 & 1 & l'\\ 0 & 0 & 0 \end{pmatrix}\begin{pmatrix} l'-1 & 1 & l' \\ m' & 0 & -m' \end{pmatrix}  \nonumber
\end{align}
where we have also used another important rule for the 3-j symbols
\begin{align}
\begin{pmatrix} l_1 & l_2 & l_3 \\ m_1 & m_2 & m_3 \end{pmatrix} =0 \quad \hbox{if} \quad m_1 + m_2 \neq -m_3\,.
\end{align}
We can then express Eq.~\eqref{e:almprime} as
\begin{align}
a'_{l'm} & =\sum_{l=1}^{+\infty} K_{l'lm} a_{lm}\,.\label{eq:boosted_alm}
\end{align}
with the kernel
\begin{align} \label{e:Kz}
K_{l'lm} = \delta_{ll'} +\delta_{l(l'+1)}A_{lm}\mathcal{D}\e{kin} + \delta_{l(l'-1)} A_{l+1m} \mathcal{D}\e{kin}\,,
\end{align}
and define the functions
\begin{align}
B_{lm}(\mathcal{D}\e{kin}/\beta) & \equiv A_{lm}  \frac{\mathcal{D}\e{kin}}{\beta}  \equiv \sqrt{\frac{l^2-m^2}{(2l+1)(2l-1)}} \cdot  \frac{\mathcal{D}\e{kin}}{\beta} 
\label{eq:A}
\end{align}
The functions $B_{lm}(\mathcal{D}\e{kin}/\beta)$ depend linearly on the ratio $\mathcal{D}\e{kin}/\beta$, while the correlation coefficients $A_{lm}$'s are independent of $\beta$ and $\mathcal{D}\e{kin}$. For $|m|=l$ these coefficients vanish while for fixed $|m|\ll l$, they are typically of order $1/2$ and they have the  large $l$ behavior
\begin{align}
\lim_{l\to +\infty}A_{lm} & = \frac{1}{2} \,.
\end{align}
We plot $A_{lm}$ in Fig.\,\ref{fig:A} for relevant values of $l$ and $m$. As one sees in the figure, the correlation coefficients are always positive and largest contributions to the cross correlation coefficients $A_{lm}$ come from the smaller values of $m$.
In the next section, we compute the correlation between different $a_{lm}'$'s, assuming that the Universe is statistically isotropic and homogeneous for the  observer $O$.


\section{Correlation of neighboring multipoles}\label{sec:Correlation}
We assume that  $O$ is a comoving observer for which the Universe appears statistically isotropic. This implies that the 2-point correlation function measured by $O$ satisfies
\begin{align}
\langle a_{lm} a_{l'm'}^* \rangle = C_l \delta_{ll'} \delta_{mm'}\,. \label{eq:statistical_isotropy}
\end{align}
Here and in what follows, brackets denote ensemble averages. For the boosted observer, $O'$, to linear order in $\beta$ we find, using \eqref{eq:boosted_alm},
\begin{widetext}
\begin{align}
\langle a'_{lm} (a'_{l'm'})^* \rangle  & = C_{l} \delta_{ll'}\delta_{mm'} + \mathcal{D}\e{kin} \cdot  \delta_{mm'}\Big(  \left[  C_l +  C_{(l-1)}\right] A_{lm}
\delta_{l(l'+1)} + \left[ C_l +  C_{(l+1)}\right] A_{l+1m}\delta_{l(l'-1)})\Big) \label{eq:Dkin_lplus1}\\
& =C_{l} \delta_{ll'}\delta_{mm'} + \beta \cdot  \delta_{mm'}\Big( \left[  C_l +  C_{(l-1)}\right] B_{lm}(\mathcal{D}\e{kin}/\beta)
\delta_{l(l'+1)} + \left[ C_l +  C_{(l+1)}\right] B_{l+1m}(\mathcal{D}\e{kin}/\beta)\delta_{l(l'-1)})\Big)\,.
\end{align}
\end{widetext}
This is no longer proportional to $\delta_{ll'}\delta_{mm'}$ but correlations between neighboring $a_{lm}'$'s appear. This implies that an observer which is moving with respect to the statistically isotropic perturbed universe observes a statistically anisotropic distribution of sources. These deviations from statistical isotropy are encoded in the neighboring multipoles. Intuitively, aberration squashes the perturbation in the direction of motion according to Eq.\,\eqref{eq:TR3}, which naturally leads to a preferred direction, breaking statistical isotropy.  These neighboring multipole correlations involving $C_l$ and $C_{l\pm 1}$ and are proportional to $\mathcal{D}\e{kin}$. This suggests that correlations  of $a_{lm}$ and $a_{l\pm1,m}$ of a sky map of number counts may be used to constrain the amplitude of $\mathcal{D}\e{kin}$ or of $\beta$, if the ratio $\mathcal{D}\e{kin}/\beta$ is known. In the following section, we lay out the procedure to estimate $\mathcal{D}\e{kin}$ or $\beta$ from an observed map $\dd N'/\dd \Omega'[\bs{\hat{n}'},S>S_*(\nu_o)]$.


\section{Quadratic Estimators}\label{sec:Estimator}
In this section, we derive a quadratic estimator for the kinematic dipole $\mathcal{D}\e{kin}$, we check that it is unbiased, and we compute its variance and signal to noise ratio. It is straightforward to extended this estimator to the redshift dependent $\mathcal{D}\e{kin}(z)$ by replacing $a_{lm}'\to a_{lm}'(z)$. For simplicity, we assume a catalog with full sky coverage. In practice, number count catalogues cover only a fraction of the sky, which induces important limitations which we discuss in Sec.\,\ref{sec:Discussion}. First one needs to determine the boosted coefficients using Eq.\,\eqref{eq:boosted_alm_measured}. We can estimate the variance $C_l$ of the $a'_{lm}$'s, with the following quadratic estimator
\begin{align}
\hat{C}_l = \frac{1}{2l+1}\sum_{m=-l}^l |a'_{lm}|^2\,,    
\end{align}
which is boost independent to linear order in $\beta$. The predictions for the correlators of $l$ and $l+1$ is
\begin{align}
\langle a'_{lm}(a'_{(l+1)m})^* \rangle & = [ C_{l+1} +  C_l] A_{l+1\,m}\mathcal{D}\e{kin} \,.
\end{align}
We can estimate $\mathcal{D}\e{kin}$ with all products of neighboring $a'_{lm}$'s available,
\begin{align}
\hat{\mathcal{D}}\e{kin} = \frac{1}{l_{\max}(l\e{max}+2)} \sum_{l=1}^{l\e{max}} \sum_{m=-l}^{l} \frac{a'_{lm}(a'_{(l+1)m})^*}{D_{lm}}\,,\label{eq:Redshift_Dipole_Estimator}
\end{align}
where 
\begin{align}
D_{lm} \equiv  (C_{l+1} + C_l) A_{l+1\,m}\,.
\end{align}
Similarly, the redshift dependent kinematic dipole $\mathcal{D}\e{kin}(z)$ can be estimated using Eq.\,\eqref{eq:Redshift_Dipole_Estimator} by replacing $a'_{lm}\to a'_{lm}(z)$ and their variance $C_l \to C_l(z)$. If the ratio $\mathcal{D}\e{kin}/\beta$, which appears in $B_{lm}(\mathcal{D}\e{kin}/\beta)$ is known, for example, from direct number count dipole measurements (together with $x$ and $\alpha$), one can directly estimate $\beta$, using the following estimator
\begin{align}
\hat{\beta} = \frac{1}{l\e{max}(l\e{max}+2)}  \sum_{l=1}^{l\e{max}} \sum_{m=-l}^{l} \frac{a'_{lm}(a'_{(l+1)m})^*}{(C_{l+1} + C_l)B_{l+1m}(\mathcal{D}\e{kin}/\beta)}\,.
\end{align}
One can easily check that these estimators are unbiased. For example, we compute the bias of $\hat{\mathcal{D}}\e{kin}$
\begin{align}
\qquad & b_{\mathcal{D}\e{kin}} (\hat{\mathcal{D}}\e{kin})  = \langle\hat{\mathcal{D}}\e{kin}\rangle - \mathcal{D}\e{kin} \\
& = \left \langle \frac{1}{ l_{\max}(l\e{max}+2)} \sum_{l=1}^{l\e{max}} \sum_{m=-l}^{l} \frac{a'_{lm}(a'_{(l+1)m})^*}{(C_{l+1} + C_l) A_{l+1m}} \right \rangle - \mathcal{D}\e{kin} \nonumber\\
& =   \frac{1}{ (l_{\max}(l\e{max}+2)} \sum_{l=1}^{l\e{max}} \sum_{m=-l}^{l} \frac{\langle a'_{lm}(a'_{(l+1)m})^*\rangle }{(C_{l+1} + C_l) A_{l+1m}}  - \mathcal{D}\e{kin} \nonumber \\
& = \l(\frac{1}{l_{\max}(l\e{max}+2)} \sum_{l=1}^{l\e{max}} \sum_{m=-l}^{l} \mathcal{D}\e{kin}\r) - \mathcal{D}\e{kin}   = 0 \,,\nonumber
\end{align}
which shows that our estimator is unbiased. Next, we compute the variance of $\hat{\mathcal{D}}\e{kin}$
\begin{align}\label{eq:variance}
\mathbb{V}\hbox{ar}(\hat{\mathcal{D}}\e{kin})= \langle \hat{\mathcal{D}}\e{kin}^2\rangle - \l[\langle\hat{\mathcal{D}}\e{kin} \rangle\r]^2 = \langle\hat{\mathcal{D}}\e{kin}^2\rangle- \mathcal{D}\e{kin}^2\,.
\end{align}
The squared expectation value of $\hat{\mathcal{D}}\e{kin}$ is given by 
\begin{widetext}
\begin{align}\label{eq:squared_Expectation}
\langle\hat{\mathcal{D}}\e{kin}\rangle^2 = \l( \frac{1}{ l_{\max}(l\e{max}+2)}\r)^2 \sum_{l=1}^{l\e{max}} \sum_{l'=1}^{l\e{max}} \sum_{m=-l}^{l}\sum_{m'=-l'}^{l'} \frac{\langle a'_{lm} (a'_{(l+1)m})^* \rangle \langle a'_{l'm'}(a'_{(l'+1)m'})^* \rangle}{D_{lm}D_{l'm'}}\,.
\end{align}
The expectation value of $\hat{\mathcal{D}}\e{kin}^2$ reads
\begin{align}
\langle\hat{\mathcal{D}}\e{kin}^2\rangle = \l( \frac{1}{ l_{\max}(l\e{max}+2)}\r)^2 \sum_{l=1}^{l\e{max}} \sum_{l'=1}^{l\e{max}} \sum_{m=-l}^{l}\sum_{m'=-l'}^{l'} \frac{\langle a'_{lm} (a'_{(l+1)m})^* a'_{l'm'}(a'_{(l'+1)m'})^* \rangle}{D_{lm}D_{l'm'}}\,.
\end{align}
We can use Isserli's theorem~\cite{Isserlis:1918} (better known as Wick's theorem~\cite{Wick:1950}) to express the expectation value of four Gaussian random  variables as a sum of products of expectation values of 2 random variables. We have
\begin{align}
\langle a'_{lm} (a'_{(l+1)m})^* a'_{l'm'}(a'_{(l'+1)m'})^* \rangle & =\langle a'_{lm} (a'_{(l+1)m})^*\rangle \langle  a'_{l'm'}(a'_{(l'+1)m'})^* \rangle \nonumber \\
& \quad + \langle a'_{lm} a'_{l'm'}\rangle \langle ( a'_{(l+1)m})^* (a'_{(l'+1)m'})^*\rangle \nonumber\\
& \quad + \langle a'_{lm} (a'_{(l'+1)m'})^* \rangle \langle (a'_{(l+1)m})^* a'_{l'm'}\rangle \nonumber \\
& = \langle a'_{lm} (a'_{(l+1)m})^*\rangle \langle  a'_{l'm'}(a'_{(l'+1)m'})^* \rangle \nonumber \\
& \quad + (-1)^{m'}(-1)^m  \langle a'_{lm} (a'_{l'(-m')})^*\rangle  \langle a'_{(l+1)(-m)} (a'_{(l'+1)m'})^*\rangle\nonumber \\
& \quad + \langle a'_{lm} (a'_{(l'+1)m'})^* \rangle \langle  a'_{l'm'} (a'_{(l+1)m})^*\rangle\,.
\end{align}
where we have used that $a_{lm}^* = (-1)^m a_{l(-m)}$. The first term on the right hand side ends up cancelling with the $\langle\hat{\mathcal{D}}\e{kin}\rangle^2$ from \eqref{eq:squared_Expectation} in \eqref{eq:variance}.  We are left with
\begin{align}
\mathbb{V}\hbox{ar}(\hat{\mathcal{D}}\e{kin}) & = \frac{1}{l^2_{\max}(l\e{max}+2)^2}   \sum_{l=1}^{l\e{max}}\sum_{l'=1}^{l\e{max}} \sum_{m=-l}^{l}\sum_{m'=-l'}^{l'} \Bigg( \frac{(-1)^{m + m'} C_{l}\delta_{ll'} \delta_{m(-m')} C_{l+1} \delta_{(l+1)(l'+1)} \delta_{(-m)m'}}{D_{lm}D_{l'm'}}  + \frac{C_{l}\delta_{l(l'+1)} \delta_{mm'} C_{l'} \delta_{l'(l+1)} \delta_{m'm}}{D_{lm}D_{l'm'}}\Bigg) \nonumber \\
& = \frac{1}{l^2_{\max}(l\e{max}+2)^2}   \sum_{l=1}^{l\e{max}}\sum_{l'=1}^{l\e{max}} \sum_{m=-l}^{l} \frac{ \delta_{ll'} C_{l}C_{l+1}} {D_{lm}D_{l'(-m)}} = \frac{1}{l^2_{\max}(l\e{max}+2)^2}   \sum_{l=1}^{l\e{max}} \sum_{m=-l}^{l} \frac{C_l C_{l+1}}{D_{lm}D_{l(-m)}}\,.
\end{align}
\end{widetext}
The signal to noise ratio reads
\begin{align}\label{eq:SNR}
\hbox{S/N} = \frac{\langle \hat{\mathcal{D}}\e{kin} \rangle }{\sqrt{\mathbb{V}\hbox{ar}(\hat{\mathcal{D}}\e{kin})}} = \frac{\mathcal{D}\e{kin} \cdot  l_{\max}(l\e{max}+2)}{\sqrt{\sum_{l=1}^{l\e{max}} \sum_{m=-l}^l \frac{C_l C_{l+1}}{D_{lm} D_{l(-m)}}}} \,.
\end{align}
Future number count experiments are expected to measure $\dd N'/\dd \Omega'[\bs{\hat{n}'},S>S_*(\nu_o)]$ with an angular resolution below the arcmin scale (with e.g.\,30 galaxies/arcmin$^2$ for Euclid \cite{Laureijs2011}), translating to $l\e{max} \ge 10^4$. It is not relevant here that the $C_l$'s cannot be calculated within linear perturbation theory. They can be extracted from the observations themselves even in the non-linear regime. The only relevant assumptions are statistical isotropy for observer $O$, as we show in App.\,\ref{app:Statistical_Isotropy} and $\beta\ll 1$ so that an expansion to linear order in $\beta$ makes sense. 
To estimate how the signal to noise ratio scales with $l\e{max}$, we estimate the denominator of \eqref{eq:SNR} for a smooth variance of the source number counts so that we can set $C_l \simeq C_{l+1}$, such that 
\begin{align}
\sum_{l=1}^{l\e{max}} \sum_{m=-l}^{l} \frac{C_l C_{l+1}}{D_{lm}D_{l(-m)}} \simeq \sum_{l=1}^{l\e{max}} \sum_{m=-l}^{l} \frac{1}{4 A_{l+1m} A_{l+1(-m)}}\,.
\end{align}
To estimate the denominator, we replace the sum over $m$ by an integral and use $A_{l(-m)} = A_{lm}$ 
\begin{align}
\sum_{m=-l}^{l} \frac{1}{4 [A_{l+1m}]^2}  & = \sum_{m=-l}^l \frac{ (2l+3)(2l+1)}{4  [(l+1)^2 - m^2]} \nonumber\\ &\simeq \frac{(2l+3)(2l+1)}{4} \int_{-l}^l \frac{\dd m}{[(l+1)^2 - m^2]} \nonumber \\
& = \frac{(2l+3)(2l+1)  \log(2l+1)}{4 (l+1)}
\end{align}
We now integrate this result over $l$ to find 
\begin{widetext}
\begin{align}
\sum_{l=1}^{l\e{max}} \sum_{m=-l}^{l} \frac{1}{4 [A_{l+1m}]^2}  & \simeq \int_{1}^{l\e{max}} \dd l \, \frac{(2l+3)(2l+1) \log(2l+1)}{4  (l+1)} \\
& = \frac{1}{8 } \Big [\l(4 l\e{max} (l\e{max} +2) - 2 \log(2l\e{max}+2)\r) \log(2l\e{max}+ 1) \nonumber\\
& \quad + 2 {\rm Li}_2(-3) - 2 {\rm{Li}}_2(-2 l\e{max}-1) + 8 - 2l\e{max}(l\e{max}+3) - 15 \log(3) + \log(4)\log(9)\Big] \\
& \leq  6 l\e{max}^2 \label{e:sumint}
\end{align}
\end{widetext}
where the validity of the last inequality can be checked numerically for $l\e{max}\leq 10^5$.
Here Li$_2$ denotes the dilogarithm given by
\be
{\rm Li}_2(z) = \int_z^0\frac{\log(1-t)}{t}\dd t = \sum_{k=1}^\infty \frac{z^k}{k^2} \,.
\ee
With \eqref{e:sumint}, we can set
\begin{empheq}[box=\fbox]{equation}
\hbox{S/N} \geq \frac{\mathcal{D}\e{kin} \cdot l\e{max}}{\sqrt{6}}
\end{empheq}
which is larger than 1 for $l\e{max}\geq 497$, assuming $\hat{\beta}$ converges to $\beta\e{dip}$ and  $\mathcal{D}\e{kin} = 4 \beta\e{dip}$. Note that this is rather a conservative estimate, since $\mathcal{D}\e{kin}$ is rather observed to be twice as large as $4\beta\e{dip}$ \cite{Secrest:2020has,Secrest:2022uvx,Bengaly:2017slg}. For future experiments like Euclid which can measure the clustering of photometric sources up to $l\e{max} = 10^4$, we can hope to constrain $\mathcal{D}\e{kin}$ (or $\beta$ if the ratio $\mathcal{D}\e{kin}/\beta$ is known) with less than 5$\%$ error. As stated earlier, it is irrelevant that the $C_l$'s cannot be calculated within linear perturbation theory, as they can be extracted from the observations themselves even in the non-linear regime. This rather suggests a clever use of the $C_l$'s beyond the linear regime.


\section{Orientation}\label{sec:orientation}
In our treatment so far, we have assumed that the direction of the peculiar velocity $\bs\beta$ is known and we have chosen the $\bs{\hat z}$-axis in its direction.  This is especially relevant, since most radio surveys agree relatively well with the direction of the CMB dipole but they find a much too large amplitude for the velocity. However, in general we want to determine both, the amplitude and the direction of $\bs\beta$.
This can be achieved easily, remembering the transformation of the coefficients $a_{lm}$ under rotation.

We start from Eq.~\eqref{eq:boosted_alm} which relates the  $a_{lm}$'s of the comoving observer to the  $a'_{lm}$'s of the observer boosted along the $\bs{\hat z}$-axis. Let us assume that the velocity $\bs\beta$ is not along the $\bs{\hat z}$ axis, but along a direction which is rotated wrt.\,$\bs{\hat z}$ by a rotation $R\in \rm{SO}(3)$. If we rotate the coordinate system by $R^{-1}=R^T$, $\bs\beta$ points in  $\bs{\hat z}$-direction wrt.\,the new coordinate system. In this rotated system,
Eq.~\eqref{eq:boosted_alm} is valid.
Under a rotation by $R^{-1}$, the $a_{lm}$'s transform as
\be
a_{lm}^{\rm rot} = \sum_{m'=-l}^l D_{mm'}^{(l)}(R)a_{lm'} \,.
\ee
Here $D_{mm'}^{(l)}(R)$ is the representation matrix of the representation $D^{(l)}$ of $\rm{SO}(3)$, see~\cite{Durrer:2020fza}, Appendix~4 for details.
In the rotated system, we have
\bea
a_{l'm}^{\rm rot \prime} &=& \sum_{m''=-l'}^{l'} D_{mm''}^{(l')}(R)a'_{l'm''} =\sum_l K_{l'lm}a_{lm}^{\rm rot} \nonumber \\
&=& \sum_{lm'} K_{l'lm}D_{mm'}^{(l)}(R)a_{lm'} \,. 
\eea
Hence, the general relation for $\bs\beta$ rotated by $R$ with respect to the $\bs{\hat z}$-axis is
\begin{widetext}
\begin{align}
a'_{l'm} &=\sum_{lm'} K^{\rm rot}_{l'lmm'} a_{lm'} \\
K^{\rm rot}_{l'lm m'} & =\sum_{m''}(D_{mm''}^{(l')}(R))^* K_{l'lm''}D_{m''m'}^{(l)}(R)\\
&= \delta_{ll'}\delta_{mm'}  + \mathcal{D}\e{kin} 
\Bigg( \delta_{l (l'+1)}  \sum_{m''}(D_{mm''}^{(l')}(R))^*A_{lm''} D_{m''m'}^{(l)}(R) + \delta_{l(l'-1)} 
\sum_{m''}(D_{mm''}^{(l')}(R))^*A_{(l+1)m''}D_{m'm''}^{(l)}(R) \Bigg)\,.
\end{align}
\end{widetext}
Here, we have used that the $D^{(l)}$ are unitary representations, $D_{mm''}^{(l')}(R^{-1}) = (D_{mm''}^{(l')}(R))^*$.

In general, a rotation is given by three Euler angles. Since $R$ is the rotation that turns $\bs{\hat{\beta}}$ with its polar angles $(\theta_\beta,\varphi_\beta)$ into the $\bs{\hat z}$-axis, we can simply choose the Euler angles $(0,-\theta_\beta,-\varphi_\beta)$ and insert
\begin{align}
D_{m''m'}^{(l')}(R)=D_{m''m'}^{(l)}(0,-\theta_\beta,-\varphi_\beta) = \sqrt{\frac{4\pi}{2l+1}}\, _{-m'}Y_{l m''}(\theta_\beta,\varphi_\beta)\,.
\end{align}
Here $_sY_{l m}$ is the spin weighted spherical harmonic of spin $s$ (see ~\cite{Durrer:2020fza}, Appendix 4 for details).
In addition to the amplitude dependence given in Eqs.~\eqref{e:Kz} to \eqref{eq:A}, the coefficients $ K^{\rm rot}_{l'lm m'}$ now also depend on the orientation $(\theta_\beta,\varphi_\beta)$. Denoting
\begin{widetext}
\bea
A^{\rm rot}(l,m,m',\theta_\beta, \varphi_\beta) &=& \sum_{m''}D_{mm''}^{(l-1) *}(0,-\theta_\beta,-\varphi_\beta)A_{lm''}
 D_{m''m'}^{(l)}(0,-\theta_\beta,-\varphi_\beta) \,, \label{eq:Rotated_Aplus}
\eea
we have
\bea
 K^{\rm rot}_{l'lm m'}(\mathcal{D}\e{kin},\theta_\beta,\varphi_\beta) &=& \delta_{ll'}\delta_{mm'}  + \delta_{l (l'+1)}A^{\rm rot}(l,m,m',\theta_\beta, \varphi_\beta) \mathcal{D}\e{kin}
+ \delta_{l (l'-1)}
A^{\rm rot}(l+1,m,m',\theta_\beta, \varphi_\beta) \mathcal{D}\e{kin} \,.
\eea 
\end{widetext}
Note that under rotations $A^{\rm rot}(l,\cdots)$ transforms like a rank $l-1$ tensor from the left and a rank $l$ tensor from the right or vice versa. (Note that since $A(l,m,m',0,0) \propto \delta_{mm'}$, an action from the left or from the right cannot be distinguished.)
For $\theta_\beta=\varphi_\beta =0$, the tensor $A^{\rm rot}$ is diagonal,
\be
A^{\rm rot}(l,m,m',0, 0) = \delta_{mm'}
A_{lm}\,.
\ee
The product $ a_{lm}a^*_{(l+1)\, m'}$ actually yields an unbiased estimator for $A^{\rm rot}(l,m,m',\theta_\beta, \varphi_\beta) \cdot \mathcal{D}\e{kin}$. To extract, the three unknowns $(\mathcal{D}\e{kin},\theta_\beta, \varphi_\beta)$, which allow to reconstruct the vector $\bs{\mathcal{D}\e{kin}} = \mathcal{D}\e{kin} \cdot \bs{\hat{\beta}}$, one computes for each $l, m$ and $m'$, the products $a_{lm} a_{(l+1)m'}^*$ from observations in some basis. This 3-dimensional array can be fitted by $A^{\rm rot}(l,m,m',\theta_\beta, \varphi_\beta) \cdot \mathcal{D}\e{kin}$ with the three parameters $\mathcal{D}\e{kin}$, $\theta_\beta$ and $\varphi_\beta$ which determine the kinematic dipole due to the observer's peculiar velocity. While the amplitude $\mathcal{D}\e{kin}$ enters only linearly, the dependence on the angles, $\theta_\beta$ and $\varphi_\beta$, is more complicated. The three parameters can be extracted e.g.\,via a Markov Chain Monte Carlo fitting procedure, assuming the $C_l$'s to be known and inserting the theoretical expressions for $A_{lm}$, which are independent of $\mathcal{D}\e{kin}$.


\section{Discussion}\label{sec:Discussion}
In this paper, we have worked out the correlation of neighboring multipoles in the number count spherical harmonic coefficients due to a boost with velocity $||\bs{\beta}||\ll 1$, in an otherwise isotropic Universe. We have found that these correlations are proportional to the kinematic dipole $\mathcal{D}\e{kin}$. We derived an unbiased estimator $\hat{\mathcal{D}}\e{kin}$ of $\mathcal{D}\e{kin}$ in terms of the observed (boosted) coefficients $a'_{lm}$, $l\in \mathbb{N}$, $m\in [-l, \dots, l]$. The same estimator with redshift dependent coefficients can be used to measure $\mathcal{D}\e{kin}(z)$, if the sources are arranged in redshift bins. Of course, the statistics for each redshift bin then decline. We computed the variance of the estimator $\hat{\mathcal{D}}\e{kin}$ and have shown that for reasonably smooth variance of the source number counts, the signal to noise ratio scales (up to $\log l\e{max}$ corrections) as $\mathcal{O} (l\e{max} \mathcal{D}\e{kin})$, which becomes larger than $1$ for $l\e{max} \geq 497$ for the expected peculiar velocity $\beta\e{dip}$ and kinematic dipole $\mathcal{D}\e{kin} \sim 4 \beta\e{dip}$. This implies that order $5\%$ precision may be achieved on $\beta$ if $l\e{max}\geq 10^4$. Note that this is a conservative estimate since the kinematic dipole is rather observed to be twice larger than $4\beta\e{dip}$ \cite{Secrest:2020has,Secrest:2022uvx,Bengaly:2017slg}. We have also derived an unbiased estimator $\hat{\beta}$ of $\beta$ which requires the additional knowledge of the ratio $\mathcal{D}\e{kin}/\beta$. The latter can be estimated via Eq.\,\eqref{eq:Redshift_Dipole} by assuming a cosmological model and by measuring $x(z)$. Alternatively, assuming that the redshift evolutions of $\alpha$ and $x$ are uncorrelated (see \cite{Dalang:2021ruy}), one can measure them from the data and use Eq.\,\eqref{eq:Ellis_Baldwin} to obtain $\mathcal{D}\e{kin}$. A last option if the redshift evolutions of $\alpha$ and $x$ are of concern, is to build a parametric model for them and fit them together with $\beta$ on tomographic measurements of $\mathcal{D}\e{kin}$ with a sufficient number of redshift bins. We studied the effects of the orientation of $\bs{\hat\beta}$ on the $l$ and $l\pm 1$ correlations and layed out the procedure which allows  to determine also the orientation of $\bs{\hat{\beta}}$ from observations. 

While our results are optimistically set in full sky, they will be affected by an incomplete sky coverage, which also breaks statistical isotropy. In practice, number count catalogues cover only a fraction of the sky, the rest being masked for a number of reasons which include but are not limited to the footprint of the detector, the milky way, obscuration by dust, turbulence from the atmosphere, bad detectors, instability of noise. One might then worry that by naively using the estimator given in \eqref{eq:Redshift_Dipole_Estimator}, one may find deviations from statistical isotropy, which actually come from the mask rather than from the peculiar velocity of the observer. One way to handle this is to account for the mask by multiplying the underlying number counts by a mask function $W(\bn)$ which  depends on the direction and on redshift and varies between zero and one, depending on the completeness of the survey. This multiplication by the mask function makes it a convolution in harmonic space, which in principle, allows to disentangle neighboring multipole correlations generated by the mask from the ones, generated by the peculiar velocity of the observer. Masking in harmonic space has been successfully carried out in the analysis of microwave background fluctuations and of galaxy number counts \cite{DES:2018csk,DES:2021dpy}.

Future work is needed to incorporate the effect of a survey mask on the estimator given in Eq.\,\eqref{eq:Redshift_Dipole_Estimator} and the impact on the estimator's bias, variance and signal to noise ratio. The current work lays the foundation to use source number counts as an assessment of how we observe the Universe moving with our special velocity, if not from a special place.


\section*{Acknowledgements}
We thank Camille Bonvin, Louis Legrand, Nastassia Grimm, Blake Sherwin and specially Azadeh Moradinezhad Dizgah for useful conversations.
This work is supported by the Swiss National Science Foundation.
F.L. acknowledges support from the Swiss National Science Foundation through grant number IZSEZ0\_207059, and from Progetto di Eccellenza 2022 of the Physics and Astronomy department at the University of Padova "Physics of the Universe", within the grant "CosmoGraLSS — Cosmology with Gravitational waves and Large Scale Structure".


\bibliographystyle{unsrt}
\bibliography{references}



\appendix

\section{Statistical Isotropy}\label{app:Statistical_Isotropy}
In this appendix, we show for completeness that Eq.\,\eqref{eq:statistical_isotropy} relies only on statistical isotropy. In particular, it does not rely on the validity of perturbation theory or on gaussianity of the fluctuations. Consider an observable, which is a real valued function, defined on the sky, i.e.\,$\mathcal{O}:\S_2 \to \mathbb{R}$. This can be the number count density or the temperature fluctuations of the CMB. Statistical isotropy implies that the correlation function depends only on the angle $\theta$ between $\bs{\hat{n}}$ and $\bs{\hat{n}'}$, 
\begin{align}
\l \langle \mathcal{O}(\bs{\hat{n}}) \mathcal{O}(\bs{\hat{n}'})\r \rangle = C(\mu)\,,\label{eq:OO_correlation}
\end{align}
where $\mu \equiv \cos(\theta) = \bs{\hat{n}}\cdot \bs{\hat{n}'}$. This function can be expanded on the interval $-1\leq \mu \leq 1$ in Legendre polynomials
\begin{align}
C(\mu) = \sum_{l=0}^{+\infty} a_l P_l(\mu) = 4\pi \sum_{l=0}^{+\infty} \frac{a_l}{2l +1} \sum_{m=-l}^l Y_{lm}(\bs{\hat{n}}) Y_{lm}^*(\bs{\hat{n}'})\label{eq:Cmu_1}
\end{align}
where we have used the addition theorem for spherical harmonics in the second equality. On the other hand, the function on the sphere $\mathcal{O}(\bs{\hat{n}})$ can be expanded in spherical harmonics
\begin{align}
\mathcal{O}(\bs{\hat{n}}) = \sum_{l=0}^{+\infty} \sum_{m=-l}^{l} a_{lm} Y_{lm}(\bs{\hat{n}}) = \sum_{l=0}^{+\infty} \sum_{m=-l}^{l} a_{lm}^* Y_{lm}^*(\bs{\hat{n}})\,,
\end{align}
where the second equality holds because $\mathcal{O}(\bs{\hat{n}})$ is real. Plugging this expansion in Eq.\,\eqref{eq:OO_correlation}, one obtains
\begin{align}
C(\mu) = \sum_{l=0}^{+\infty}\sum_{l'=0}^{+\infty} \sum_{m=-l}^l \sum_{m'=-l'}^{l'} \langle a_{lm} a_{l'm'} \rangle Y_{lm}(\bs{\hat{n}}) Y_{l'm'}(\bs{\hat{n}'})\,.\label{eq:Cmu_2}
\end{align}
As the functions $Y_{lm}(\bs{\hat{n}}) Y_{l'm'}(\bs{\hat{n}'})$ form an orthonormal basis on the Hilbert space of square integrable function on the cross product of $2$-spheres, $L^2(\S_2 \times \S_2)$, one can identify the coefficients in \eqref{eq:Cmu_1} and \eqref{eq:Cmu_2} to conclude that
\begin{align}
\langle a_{lm} a_{l'm'}^*\rangle = \delta_{ll'}\delta_{mm'} \frac{4\pi a_l}{2l +1} \equiv \delta_{ll'}\delta_{mm'}C_l\,.
\end{align}

\end{document}